\documentclass[11pt]{article}
\usepackage{times}
\font\tt=cmtt10 at 12pt
\usepackage{subfigure}
\usepackage{amsmath}
\usepackage{array}
\usepackage{latexsym}
\usepackage{graphicx}
\def\amper{\hbox{$\leadsto$}}
\def\deltap{\hbox{$\Delta_{p}$}}
\def\outputp{\hbox{$\hbox{\it output}_{p}$}}
\def\UNDSCR{\_}

\newcommand{\FT}{{Fault\kern1pt-\kern-2pt Tolerant}}
\usepackage{subfigure}

\def\taskAtimeout{\hbox{$t_{\hbox{\small task1}}$}}
\def\taskBtimeout{\hbox{$t_{\hbox{\small task2}}$}}
\def\tIAS{\hbox{$t_{\hbox{\tiny IA\_SET}}$}}
\def\tIAC{\hbox{$t_{\hbox{\tiny IA\_CLR}}$}}
\def\mIAS{\hbox{$m_{\hbox{\tiny IA\_SET\_ALARM}}$}}
\def\mIAC{\hbox{$m_{\hbox{\tiny IA\_CLR\_ALARM}}$}}
\def\dIAS{\hbox{$d_{\hbox{\tiny IA\_SET}}$}}
\def\dIAC{\hbox{$d_{\hbox{\tiny IA\_CLR}}$}}
\def\tMIAM{\hbox{$t_{\hbox{\tiny MIA\_A}}$}}
\def\tMIAB{\hbox{$t_{\hbox{\tiny MIA\_B}}$}}
\def\tTAIAM{\hbox{$t_{\hbox{\tiny TAIA\_A}}$}}
\def\tTAIAB{\hbox{$t_{\hbox{\tiny TAIA\_B}}$}}
\def\tTEIFM{\hbox{$t_{\hbox{\tiny TEIF\_A}}$}}
\def\tTEIFB{\hbox{$t_{\hbox{\tiny TEIF\_B}}$}}
\def\dMIAM{\hbox{$d_{\hbox{\tiny MIA\_A}}$}}
\def\dMIAB{\hbox{$d_{\hbox{\tiny MIA\_B}}$}}
\def\dTAIAM{\hbox{$d_{\hbox{\tiny TAIA\_A}}$}}
\def\dTAIAB{\hbox{$d_{\hbox{\tiny TAIA\_B}}$}}
\def\dTEIFM{\hbox{$d_{\hbox{\tiny TEIF\_A}}$}}
\def\dTEIFB{\hbox{$d_{\hbox{\tiny TEIF\_B}}$}}
\def\mMIAM{\hbox{$m_{\hbox{\tiny MIA\_A\_ALARM}}$}}
\def\mMIAB{\hbox{$m_{\hbox{\tiny MIA\_B\_ALARM}}$}}
\def\mTAIAM{\hbox{$m_{\hbox{\tiny TAIA\_A\_ALARM}}$}}
\def\mTAIAB{\hbox{$m_{\hbox{\tiny TAIA\_B\_ALARM}}$}}
\def\mTEIFM{\hbox{$m_{\hbox{\tiny TEIF\_A\_ALARM}}$}}
\def\mTEIFB{\hbox{$m_{\hbox{\tiny TEIF\_B\_ALARM}}$}}
\def\mTAIA{\hbox{$m_{\hbox{\tiny TAIA}}$}}
\def\mMIA{\hbox{$m_{\hbox{\tiny MIA}}$}}
\def\mTEIF{\hbox{$m_{\hbox{\tiny TEIF}}$}}
\def\UNDSCR{\_}

\newcommand{\DIRM}{\hbox{DIR\hskip-1pt-$\hskip-2pt\mathcal M$}}
\newcommand{\DIRB}{\hbox{DIR\hskip-1pt-$\hskip-1pt\mathcal B$}}
\newcommand{\DIRA}{\hbox{DIR\hskip-1pt-$\hskip-2pt\mathcal A$}}
\newtheorem{obs}{Observation}{\bf}{\rm}

\begin{document}
\title{Design Tool To Express Failure Detection Protocols}
%
\author{Vincenzo De Florio and Chris Blondia \vspace*{3pt} \\
  University of Antwerp, Department of Mathematics and Computer Science\\
  Performance Analysis of Telecommunication Systems group\\
  Middelheimlaan 1, 2020 Antwerp, Belgium \vspace*{3pt} \\
  Interdisciplinary institute for BroadBand Technology\\
  Gaston Crommenlaan 8, 9050 Ghent-Ledeberg, Belgium}

\maketitle

%
\begin{abstract}
Failure detection protocols---a fundamental building block for crafting 
fault-tolerant distributed systems---are in many cases
described by their authors making use 
of informal pseudo-codes of their conception. Often these pseudo-codes
use syntactical constructs that are not available in
COTS programming languages such as C or C++. This translates into
informal descriptions that call for ad hoc interpretations and implementations.
Being informal, these descriptions cannot be tested by their authors, which
may translate into insufficiently detailed or even faulty specifications.
This paper tackles this problem introducing a formal syntax 
for those constructs and a C library that implements them---a tool-set
to express and reason about failure detection protocols.
The resulting specifications are longer but non ambiguous, and
eligible for becoming a standard form.
\end{abstract}

\section{Introduction}
Failure
detection constitutes a fundamental building block for crafting fault-tolerant distributed systems, and
many researchers have devoted their efforts on this direction during the last decade.
Failure detection protocols are often described by their authors making use of informal
pseudo-codes of their conception. Often these pseudo-codes use syntactical constructs
such as \textsf{repeat periodically}~\cite{ChTo43,Aguilera99,BMS02}, 
\textsf{at time $t$ send heartbeat}~\cite{Chen02,BMS02},
\textsf{at time $t$ check whether message has arrived}~\cite{Chen02},
or \textsf{upon receive}~\cite{Aguilera99}, together with several
variants (see Table~\ref{t:syntax}).
We observe that such syntactical constructs are not often found in
COTS programming languages such as C or C++, which brings
to the problem of translating the protocol specifications into running software prototypes using
one such standard language.
Furthermore the lack of a formal, well-defined, and standard
form to express failure detection protocols often leads their authors to insufficiently detailed
descriptions. Those informal descriptions in turn complicate the reading process and
exacerbate the work of the implementers, which becomes time-consuming, error-prone
and at times frustrating.

\begin{table*}
{\hbox{\begin{tabular}{|l|l|l|l|l|l|l|l|}
\hline
Construct   &NFD-E~\cite{Chen02}       &$\varphi$~\cite{Haya04t}& FD~\cite{BMS02} 
            &GMFD~\cite{RaTr99}        &$\mathcal{D}\in\diamondsuit\mathcal{P}$~\cite{ChTo43}
            &$\mathcal{HB}$~\cite{Aguilera99}&$\mathcal{HB}$-pt~\cite{Aguilera99}\\ \hline
Repeat      &            no            &           no           &     yes
            &            no            &           yes
            &            yes           &           yes              \\
periodically&                          &                        &
            &                          &
            &                          &                            \\
Upon $t =$  &            yes           &           no           &     yes
            &            yes           &           no
            &            no            &           no               \\
current time&                          &                        &
            &                          &
            &                          &                            \\
Upon receive&            yes           &           yes          &     yes
            &            yes           &           yes
            &            yes           &           yes              \\
message     &                          &                        &
            &                          &
            &                          &                            \\
Concurrency &            yes           &           yes          &     no
            &            no            &           yes
            &            yes           &           yes              \\
management  &                          &                        &
            &                          &
            &                          &                            \\ \hline
\end{tabular}}}
\caption{Syntactical constructs used in several failure detector protocols.
   $\varphi$ is the accrual failure detector~\cite{Haya04t}.
   $\mathcal{D}$ is the eventually perfect failure detector of~\cite{ChTo43}.
   $\mathcal{HB}$ is the Heartbeat detector~\cite{Aguilera99}.
   $\mathcal{HB}$-pt is the partition-tolerant version of the Heartbeat detector.
  By ``Concurrency management'' we mean coroutines, threading or forking.\label{t:syntax}}
\end{table*}

Several researchers and practitioners are currently arguing that failure detection
should be made available as a network service~\cite{Haya04,Vanrenesse98gossipstyle}.
To the best of our knowledge no such service exists to date. Lacking such tool, it is important to devise methods
to express in the application layer of our software even the most complex failure
detection protocols in a simple and natural way.

In the following we introduce one such method---a class of ``time-outs'',
i.e., objects that postpone a certain function call by a
given amount of time. This feature converts time-based events
into non time-based events such as message arrivals and easily expresses
the constructs in Table~\ref{t:syntax} in standard C. In some cases,
our class removes the need of concurrency management requirements
such as coroutines or thread management libraries.
The formal character of our method allows rapid-prototyping of the algorithms
with minimal effort. This is proved through a 
Literate Programming~\cite{Knuth92} framework that produces from
a same source file both the description meant for dissemination
and a software skeleton to be compiled in standard C or C++.

The rest of this article is structured as follows: 
Section~\ref{s:tool} introduces our tool.
In Sect.~\ref{s:cases} we use it to express three classical
failure detectors.
Section~\ref{s:bb} is a case study describing a software system built with our tool.
Our conclusions are drawn in Sect.~\ref{s:end}.

\section{Time-out Management System}\label{s:tool}
This section briefly describes the architecture
of our time-out management system (TOM). 
%
The TOM class appears to the user as a couple of new types
and a library of functions. Table~\ref{examp} provides an idea of the
client-side protocol of our tool.

\begin{table*}
{\begin{tabular}{ll}
{\bf 1.}&   /* declarations */\\
        &  TOM *tom;\\
        &  timeout\UNDSCR{}t t1, t2, t3;\\
        &  int my\UNDSCR{}alarm(TOM*), another\UNDSCR{}alarm(TOM*);\\
{\bf 2.}&  /* definitions */\\
        &  tom $\leftarrow$ tom\UNDSCR{}init(my\UNDSCR{}alarm);\\
        &  tom\UNDSCR{}declare(\&t1, TOM\UNDSCR{}CYCLIC, TOM\UNDSCR{}SET\UNDSCR{}ENABLE, 
        TIMEOUT1, SUBID1, DEADLINE1);  \\
        &  tom\UNDSCR{}declare(\&t2, TOM\UNDSCR{}NON\UNDSCR{}CYCLIC, TOM\UNDSCR{}SET\UNDSCR{}ENABLE, 
         TIMEOUT2, SUBID2, DEADLINE2);  \\
        &  tom\UNDSCR{}declare(\&t3, TOM\UNDSCR{}CYCLIC, TOM\UNDSCR{}SET\UNDSCR{}DISABLE, 
        TIMEOUT3, SUBID3, DEADLINE3);  \\
        &  tom\UNDSCR{}set\UNDSCR{}action(\&t3, another\UNDSCR{}alarm); \\
{\bf 3.}&  /* insertion */\\
        &  tom\UNDSCR{}insert(tom, \&t1), 
         tom\UNDSCR{}insert(tom, \&t2),   
         tom\UNDSCR{}insert(tom, \&t3); \\
{\bf 4.}&  /* control */\\
        &  tom\UNDSCR{}enable(tom, \&t3); \\
        &  tom\UNDSCR{}set\UNDSCR{}deadline(\&t2, NEW\UNDSCR{}DEADLINE2); \\
        &  tom\UNDSCR{}renew(tom, \&t2); \\
        &  tom\UNDSCR{}delete(tom, \&t1); \\
{\bf 5.}&  /* deactivation */\\
        &  tom\UNDSCR{}close(tom);
\end{tabular}}
\caption{Usage of the TOM class. In {\bf 1.} a time-out
list pointer and three time-out objects are declared, together with two
alarm functions. In {\bf 2.} the time-out list and the time-outs
are initialized, and a new alarm is
associated to time-out {\tt t3}. Insertion is carried out at
point {\bf 3.} At {\bf 4.}
{\tt t3} is enabled and a new deadline
value is specified for {\tt t2}. The latter is renewed
and {\tt t1} is deleted.
The list is finally deactivated in {\bf 5.}\label{examp}}
\end{table*}

To declare a time-out manager,
the user needs to define a pointer to a {\tt TOM} object and then
call function {\tt tom\UNDSCR{}init}. Argument to this function is
an alarm, i.e., the function to be called when a
time-out expires:
\begin{center}
\begin{tabular}{l}
{\tt int alarm(TOM *);} \ \ 
{\tt tom = tom\UNDSCR{}init( alarm );} 
\end{tabular}
\end{center}
The first time function {\tt tom\UNDSCR{}init} is called a custom thread is
spawned. That thread is the actual time-out manager.

Now it is possible to define time-outs.
This is done via
type {\tt timeout\UNDSCR{}t} and function {\tt tom\UNDSCR{}declare}; an example follows:
\begin{center}
{\tt timeout\UNDSCR{}t}~{\tt t}; \ \ 
{\tt tom\UNDSCR{}declare(\&t,}{\tt TOM\UNDSCR{}CYCLIC, TOM\UNDSCR{}SET\UNDSCR{}ENABLE,} 
                          {\tt TID, TSUBID, DEADLINE)}.
\end{center}
\noindent
In the above, time-out {\tt t} is declared as:
\begin{itemize}
\item A cyclic time-out (renewed on expiration; as opposed to 
{\tt TOM\UNDSCR{}NON\UNDSCR{}CYCLIC}, which means ``removed on expiration''),
\item enabled (only enabled time-outs ``fire'', i.e., call their
alarm on expiration; an alarm is disabled with {\tt TOM\UNDSCR{}SET\UNDSCR{}DISABLE}),
\item with a deadline of {\tt DEADLINE} local clock ticks before expiration.
\end{itemize}

A time-out {\tt t} is identified as a couple of integers---{\tt TID} and {\tt TSUBID}
in the above example. This is done because
in our experience it is often useful to distinguish instances of \emph{classes\/}
of time-outs. We use then {\tt TID} for the class identifier and {\tt TSUBID}
for the particular instance. A practical example of this is given in Sect.~\ref{s:bb}.

Once defined, a time-out can be submitted to the time-out manager
for insertion in its running list of time-outs---see~\cite{DeFl06} for
further details on this.
From the user point of view, this is managed
by calling function
\begin{center}
{\tt tom\UNDSCR{}insert( TOM *, timeout\UNDSCR{}t * )}.
\end{center}
Note that a time-out might be submitted to more than one time-out manager.

After successful insertion 
an enabled time-out will trigger the call of the default alarm
function after the specified deadline. If that time-out is declared as
{\tt TOM\UNDSCR{}CYCLIC} the time-out would then be re-inserted.

Other control functions are available: a time-out can be temporarily suspended
while in the time-out list via function
\begin{center}
{\tt tom\UNDSCR{}disable( TOM *, timeout\UNDSCR{}t * )}
\end{center}
\noindent
and (re-)enabled via function
\begin{center}
{\tt tom\UNDSCR{}enable( TOM *, timeout\UNDSCR{}t * )}.
\end{center}

Furthermore, the user can specify a new alarm function via
{\tt tom\UNDSCR{}set\UNDSCR{}action}) and a
new deadline via {\tt tom\UNDSCR{}set\UNDSCR{}deadline}; can
delete a time-out from the list via
\begin{center}
{\tt tom\UNDSCR{}delete( TOM *, timeout\UNDSCR{}t * )},
\end{center}
and renew\footnote{Renewing a time-out means removing and re-inserting it.} it via
\begin{center}
{\tt tom\UNDSCR{}renew( TOM *, timeout\UNDSCR{}t * )}.
\end{center}

Finally, when the time-out management service is no longer needed, the
user should call function
\begin{center}
{\tt tom\UNDSCR{}close( TOM * )},
\end{center}
\noindent
which also halts the time-out manager thread should no other client
be still active.

\subsection{System assumptions, building blocks, and algorithms}
This section is to provide the reader with a clear definition of 
\begin{itemize}
\item the system assumptions our tool builds upon,
\item the architectural building blocks of our system,
\item the algorithms managing the list of time-outs.
\end{itemize}

\subsubsection{System assumptions}
Our tool is built in C for a generic Unix-like system with threads and 
standard inter-process communication facilities. Two implementation exists to date---one
based on Embedded Parix~\cite{Anon96b}, the other using the standard Posix threads library~\cite{Pthreads}.
A fundamental requirement of our model is that processes must have access
to some local physical clock giving them the ability to measure time.
The availability of means to control the priorities of processes is also
an important factor to reducing the chances of late alarm execution.
We also assume that the alarm functions are small grained both in CPU and I/O
usage so as not to interfere ``too much'' with the tasks of the TOM.
Finally, we assume the availability of asynchronous, non-blocking primitives to
send and receive messages. 

\subsubsection{Architectural building blocks}
\begin{figure}[h]
\centerline{\includegraphics[width=0.7\textwidth]{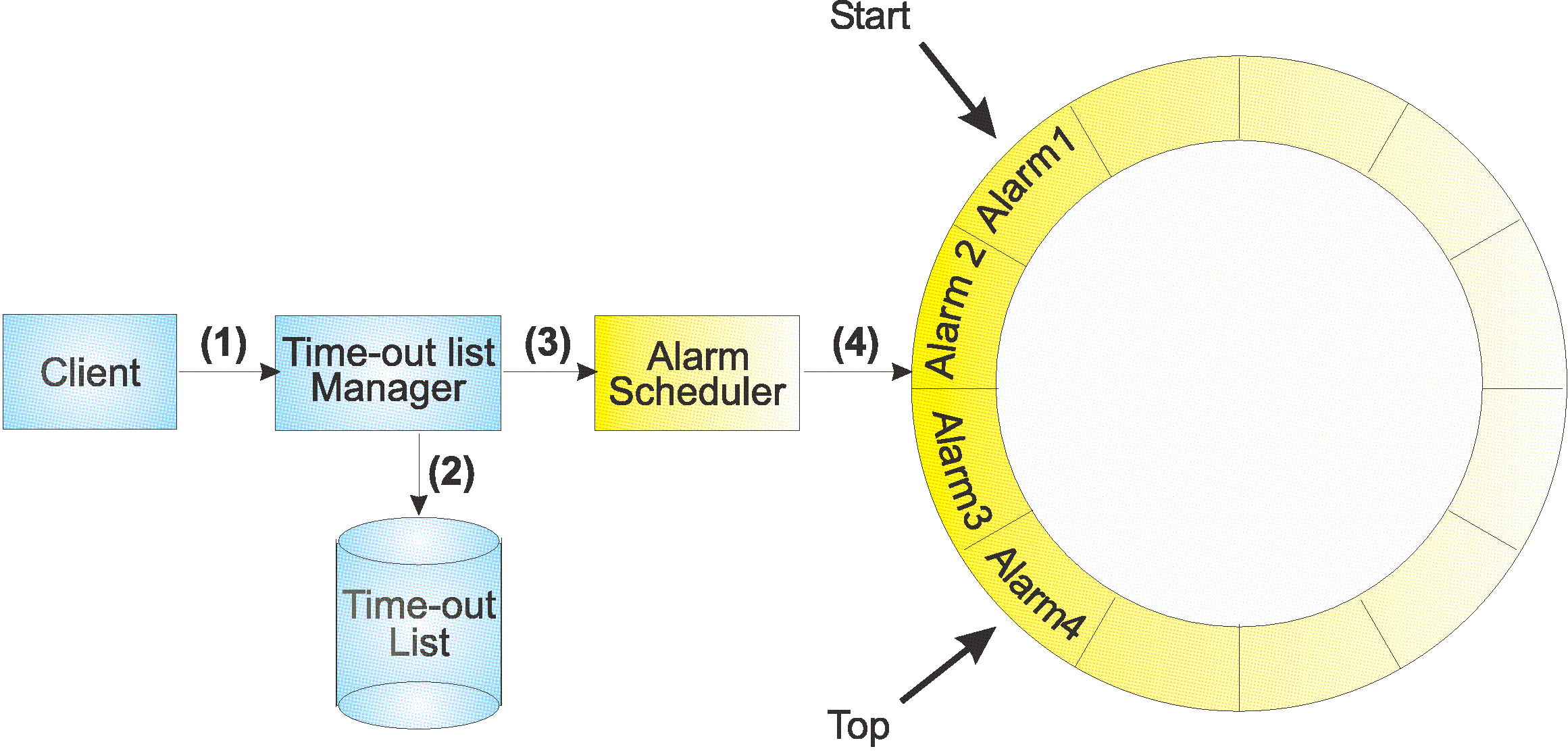}}
\caption{Architecture of the time-out management system.}
\label{buffer}
\end{figure}

Figure~\ref{buffer} portrays the architecture of our time-outs manager: in

\begin{description}
\item{(1),} the client process sends requests
to the time-out list manager; in
\item{(2),} the time-out list manager
accordingly updates the time-out list with the server-side protocol described
in Sect.~\ref{ss}.
\item{(3)} Each time a time-out reaches its deadline,
a request for execution of the corresponding alarm is sent to a task called alarm scheduler.
\item{(4)} The alarm scheduler allocates an alarm request to the first available process out of those 
in a circular list of alarm processes, possibly waiting until one of them becomes available.
\end{description}

\begin{figure}[h]
\centerline{\includegraphics[width=0.9\textwidth]{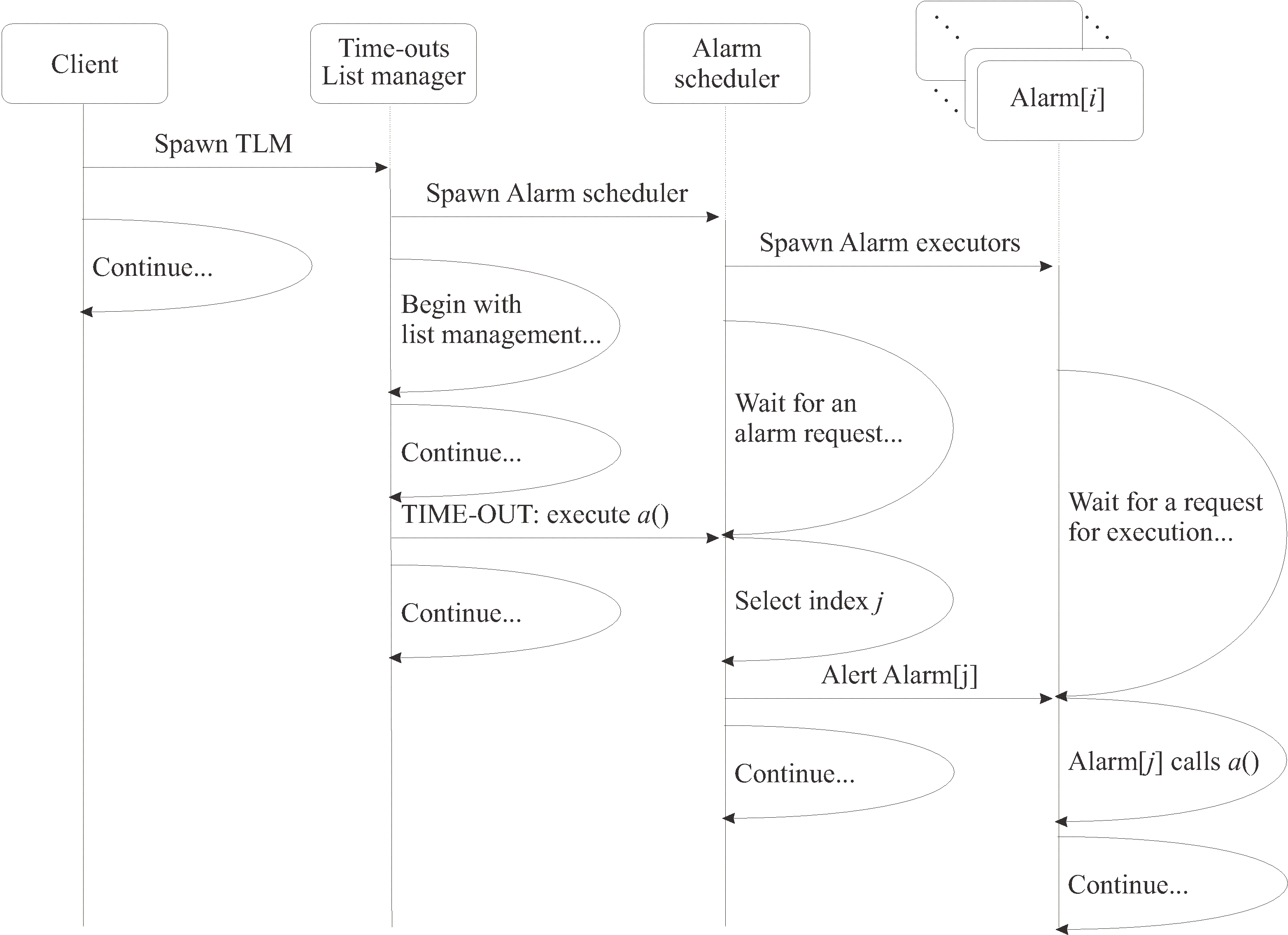}}
\caption{Sequence diagram for the tasks of the time-outs manager.}
\label{sequence}
\end{figure}

Figure~\ref{sequence} shows the sequence diagram corresponding to the initialization of the
system and the management of the first time-out request.

The presence of an alarm scheduler and of the circular list of alarm processes 
can have great consequences on performance and on the ability of our
system to fulfil real-time requirements. Such aspects
have been studied in~\cite{DeFl06}. Our system may also operate in a
simpler mode, without the above mentioned two components
and with the time-out list manager taking care of the execution of the alarms.

\subsubsection{Algorithms}\label{ss}

The server-side protocol is run by a component called time-out list
manager (TLM). 
The TLM implements a well-known
time-out queuing strategy that is described e.g. in~\cite{Tan96}.
TLM basically checks every {\sf TM\UNDSCR{}CYCLE} for the occurrence of
one of these two events:
\begin{itemize}
\item A request from a client has arrived. If so, TLM serves that request.
\item One or more time-outs have expired. If so, TLM executes
      the corresponding alarms.
\end{itemize}

Each time-out {\sf t} is characterized by its \emph{deadline}
{\sf t.deadline}, a positive integer representing the number of
clock units that must separate the time of insertion or renewal from the
scheduled time of alarm execution. This field can only be set by functions
{\sf tom\UNDSCR{}declare} and {\sf tom\UNDSCR{}set\UNDSCR{}deadline}.
Each time-out {\sf t} holds also a field, {\sf t.running},
initially set to {\sf t.deadline}.

Each time-out list object, say {\sf tom}, hosts a variable representing
the origin of the time axis. This variable, {\sf tom.start\UNDSCR{}time},
regards in particular the time-out at the top of the time-out list---the idea 
is that the top of the list is the only entry whose {\sf running}
field needs to be compared with current time in order to verify
the occurrence of the time-out-expired event. 
For the time-outs behind the top one, that field represents relative
values, viz., distances from expiration time of the closest,
preceding time-out.
In other words, the overall time-out list management aims at isolating a
``closest to expiration'' time-out, or head time-out, that is the one
and only time-out to be tracked for expiration, and at keeping
track of a list of ``relative time-outs.''

Let us call {\sf TimeNow} the system function returning the current
value of the clock register. In an ordered, coherent time-out list,
residual time \emph{for the head time-out\/} {\sf t} is given by
\begin{equation}
\hbox{\sf t.running} - (\hbox{\sf TimeNow} - \hbox{\sf tom.start\UNDSCR{}time}),
\label{eq:1}
\end{equation}
\noindent
that is, residual time minus time already passed by. Let us call
quantity~(\ref{eq:1})
as $r_1$, or head residual.
For time-out $n$, $n>1$, that is for the time-out located $n-1$
entries ``after'' the top block, let us define
\begin{equation}
r_n = r_1  + \sum_{i=2}^n \hbox{\sf t}_i\hbox{\sf .running}
\label{eq:2}
\end{equation}
\noindent
as the $n$-th residual, or residual time for time-out at entry $n$.
If there are $m$ entries in the time-out list, let us define
$r_j=0$ for any $j>m$.

It is now possible to formally define the key operations
on a time-out list: insertion and deletion of an entry.

\paragraph{Insertion}
Three cases are possible, namely insertion on top,
in the middle, and at the end of the list.
\begin{description}
\item{Insertion on top.} In this case we need to insert
a new time-out object, say $t$, such that $t\hbox{\sf .deadline} < r_1$,
or whose deadline is less than the head residual. Let us call $u$
the current top of the list. Then the following operations need to
be carried out:
\[
\left\{
\begin{array}{lll}
t\hbox{\sf .running}&\leftarrow& t\hbox{\sf .deadline} + 
                     \hbox{\sf TimeNow} - \hbox{\sf tom.start\UNDSCR{}time}
                     \\
u\hbox{\sf .running}&\leftarrow& r_1 - t\hbox{\sf .deadline}.
                    \label{ins:2}
\end{array}
\right.
\]

Note that the first operation is needed in order to verify relation
\[
t\hbox{\sf .running} - (\hbox{\sf TimeNow} - 
                        \hbox{\sf tom.start\UNDSCR{}time}) =
t\hbox{\sf .deadline},
\]
\noindent
while the second operation aims at turning the absolute value
kept in the {\sf running} field of the ``old'' head of the list into
a value relative to the one stored in the corresponding field
of the ``new'' top of the list.

\item{Insertion in the middle.} In this case we need to insert
a time-out $t$ such that
\[ \exists \, j : r_j \leq t\hbox{\sf .deadline} < r_{j+1}. \]
Let us call $u$ time-out $j+1$.
(Note that both $t$ and $u$ exist by hypothesis).
Then the following operations need to be carried out:
\[
\left\{
\begin{array}{ll}
t\hbox{\sf .running}&\leftarrow t\hbox{\sf .deadline} - r_j 
                     \\
u\hbox{\sf .running}&\leftarrow u\hbox{\sf .running} - t\hbox{\sf .running}.
                    \label{ins:4}
\end{array}
\right.
\]

\begin{obs}
Note how, both in the case of insertion on top and in that of
insertion in the middle of the list, time interval $[0,r_m]$
has not changed its length---only, it has been further subdivided,
and is now to be referred to as $[0,r_{m+1}]$.
\end{obs}

\item{Insertion at the end.} Let us suppose 
the time-out list consists of $m>0$ items, and that we need to
insert time-out $t$ such that $t\hbox{\sf .deadline}\ge r_m$.
In this case we simply tail the item and initialize it so that
\[
t\hbox{\sf .running} \leftarrow t\hbox{\sf .deadline} - r_m.
\]
\end{description}

\begin{obs}
Note how insertion at the end of the list is the only way to
prolong the range of action from a certain $[0, r_m]$ to
a larger $[0, r_{m+1}]$.
\end{obs}

\paragraph{Deletion}
The other basic management operation on the time-out list is deletion.
As we had three possible insertions, likewise
we distinguish here deletion from top,
from the middle, and from the end of the list.

\begin{description}
\item{Deletion from top.} If the list is a singleton we are
in a trivial case. Let us suppose there are at least two items in the
list.  Let us call $t$ the top of the list and $u$ the next element---the
one that will be promoted to top of the list. From its definition
we know that
\begin{eqnarray}
r_2 & = & u\hbox{\sf .running} + r_1 \nonumber\\
    & = & u\hbox{\sf .running} + t\hbox{\sf .running} - (\hbox{\sf TimeNow} - \hbox{\sf tom.start\UNDSCR{}time}).\label{del:1}
\end{eqnarray}

By (\ref{eq:1}), the bracketed quantity is the elapsed time. Then the amount
of absolute time units that separate current time from the expiration time
is given by $u\hbox{\sf .running} + t\hbox{\sf .running}$.
In order to ``behead'' the list we therefore need to update $t$ as follows:
\[
 u\hbox{\sf .running} \leftarrow u\hbox{\sf .running} + t\hbox{\sf .running}.
\]

\item{Deletion from the middle.} Let us say we have two consecutive
time-outs in our list, $t$ followed by $u$, such that $t$ is not the top
of the list. With a reasoning similar to the one just followed we get to
the same conclusion---before physically purging $t$ off the list
we need to perform the following step:
\[
 u\hbox{\sf .running} \leftarrow u\hbox{\sf .running} + t\hbox{\sf .running}.
\]

\item{Deletion from the end.} Deletion from the end means deleting
an entry which is not referenced by any further item in the list.
Physical deletion can be performed with no need for updating.
Only, the interval of action is shortened.
\end{description}

\begin{obs}
Variable {\sf tom.start\UNDSCR{}time} is never set when deleting
from or inserting entries into a time-out list, except when inserting
the first element: in such case, that variable is set to the current value 
of {\sf TimeNow}.
\end{obs}

\begin{figure}[h]
\centerline{\includegraphics[width=0.4\textwidth]{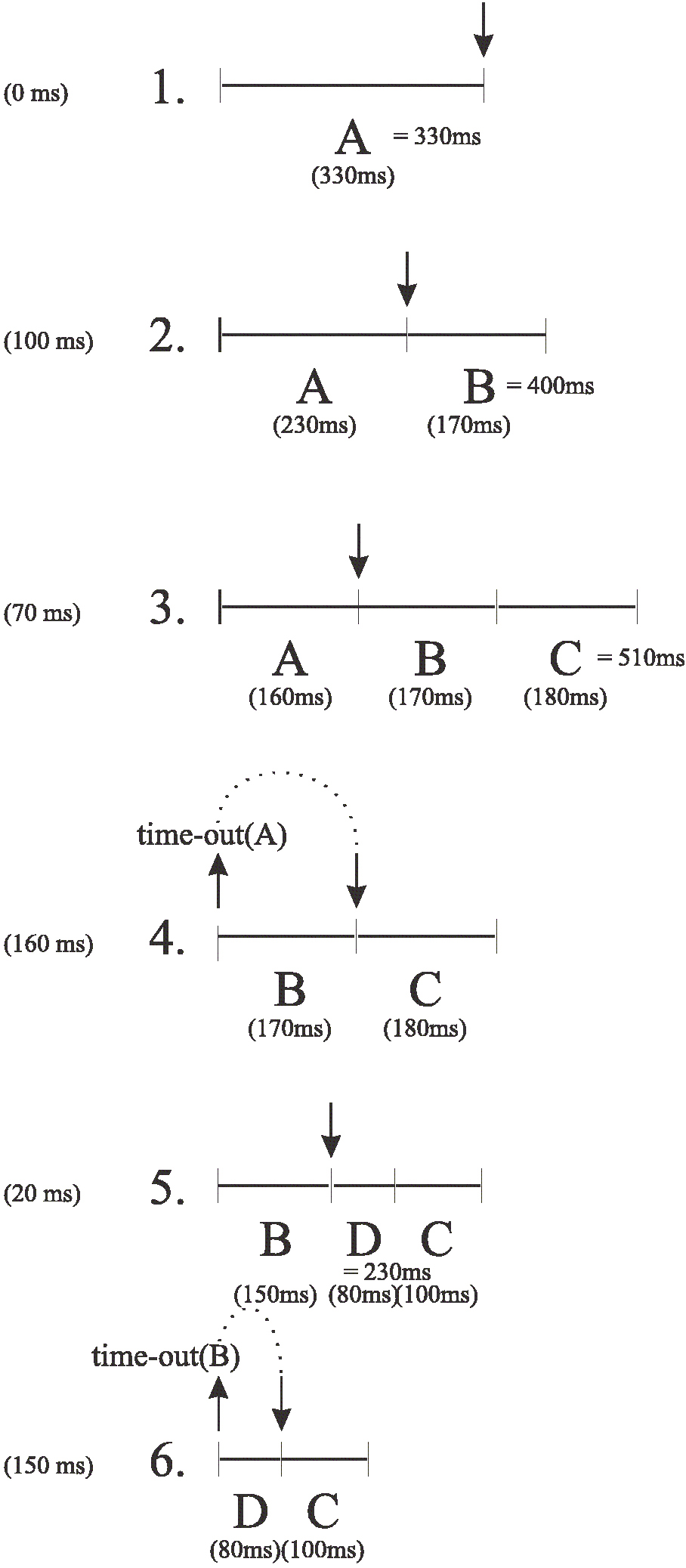}}
\caption{Operating scenario of the time-out manager.}
\label{tom3}
\end{figure}

Figure~\ref{tom3} shows the action of the server-side protocol:
In \textbf{1.}, a 330ms time-out
called \textbf{A} is inserted in the list. In \textbf{2.}, after 100ms,
\textbf{A} has been reduced to 230ms and a 400ms time-out, called \textbf{B},
is inserted (its value is 170ms, i.e., 400-230ms). Another 70ms have passed in \textbf{3.}, so
\textbf{A} has been reduced to 160ms. At that point, a 510ms time-out,
\textbf{C} is inserted and goes at the third position. In \textbf{4.}, after 160ms,
time-out \textbf{A} occurs---\textbf{B} becomes then the top of the list;
its decrementation starts. In \textbf{5.} another 20ms have passed and \textbf{B} is
at 150ms---at that point a 230ms time-out, called \textbf{D} is inserted.
Its position is in between \textbf{B} and \textbf{C}, therefore this latter is
adjusted. In \textbf{6.}, after 150ms, \textbf{B} occurs and \textbf{D} goes on top.


\section{Discussion}\label{s:cases}
In this section we show that the syntactical constructs in Table~\ref{t:syntax}
can be expressed in terms of our class of time-outs. We do so by considering three
classical failure detectors and providing their time-out based specifications.

Let us consider the classical formulation of eventually perfect
failure detector $\mathcal{D}$~\cite{ChTo43}. 
The main idea of the protocol is to require each task to send a ``heartbeat''
to its fellows and maintain a list of tasks suspected to have failed. A task identifier $q$ enters 
the list of task $p$ 
if no heartbeat is received by $p$ during a certain amount of time, $\deltap(q)$, initially
set to a default value. This value is increased when late heartbeats are received.

The basic structure of $\mathcal{D}$ is that of a
coroutine with three concurrent processes, two of which execute a task
periodically while the third one is triggered by the arrival of a message:

\begin{tabbing}
aa\=aa\=aa\=aa\=aa\=aa\=aa \kill
        \> \emph{Every process $p$ executes the following}:\\
        \\
 \> $\outputp \leftarrow 0$\\
 \> {\bf for} all $q\in\Pi$\\
	\> \> $\deltap(q)\leftarrow$ default time interval\\
        \\
	\> {\bf cobegin}\\
 \> \> || \emph{Task 1:} {\bf repeat periodically}\\
        \> \> \> send ``$p$-is-alive'' to all\\
        \\
 \> \> || \emph{Task 2:} {\bf repeat periodically}\\
        \> \> \> {\bf for} all $q\in\Pi$\\
	\> \> \> \> {\bf if} $q\not\in\outputp$ and $p$ did not receive ``$q$-is-alive'' during\\
	\> \> \> \> \> the last $\deltap(q)$ ticks of $p$'s clock {\bf then}\\
 \> \> \> \> \> \> $\outputp \leftarrow \outputp \cup \{q\}$\\
	\\
 \> \> || \emph{Task 3:} {\bf when} received ``q-is-alive'' for some $q$\\
	\> \> \> {\bf if} $q\in\outputp$\\
	\> \> \> \> $\outputp \leftarrow \outputp - \{q\}$\\
	\> \> \> \> $\deltap(q)\leftarrow \deltap(q) + 1$\\
	\> {\bf coend}.
\end{tabbing}

We call the {\bf repeat periodically} in \emph{Task 1} a ``multiplicity 1'' repeat, because
indeed a single action (sending a ``$p$-is-alive'' message) has to be tracked, while
we call ``multiplicity $q$'' repeat the one in \emph{Task 2}, which requires to check $q$
events.

Our reformulation of the above code is as follows:

\begin{tt}
\begin{tabbing}
aa\=aa\=aa\=aa\=aa\=aa \kill
\> \mbox{\emph{Every process $p$ executes the following}:}\\
        \\
 \> {\bf timeout\_t} \taskAtimeout, \taskBtimeout[NPROCS];\\
 \> {\bf task\_t} $p$, $q$;\\
 \> {\bf for} ($q$=0; $q$<NPROCS; $q$++) \{\\
 \> \> $\deltap[q]$ = DEFAULT\_TIMEOUT;\\
 \> \> $\outputp[q]$ = TRUST;\\
 \> \}\\
	\\
 \> /* \emph{``\amper'' is our symbol for the ``address-of'' operator} */\\
 \> tom\_declare(\amper{}\taskAtimeout, TOM\_CYCLIC, TOM\_SET\_ENABLE, $p$, 0, $\deltap[q]$);\\
 \> tom\_set\_action(\amper{}\taskAtimeout, action\_Repeat\_Task1);\\
 \> tom\_insert(\amper{}\taskAtimeout);\\
	\\
 \> {\bf for} ($q$=0; $q$<NPROCS; $q$++) \{\\
 \> \> {\bf if} ($p \neq q$) \{\\
 \> \> \> tom\_declare(\taskBtimeout$+q$, TOM\_CYCLIC, TOM\_SET\_ENABLE, $q$, 0, $\deltap[q]$);\\
 \> \> \> tom\_set\_action(\taskBtimeout$+q$, action\_Repeat\_Task2);\\
 \> \> \> tom\_insert(\amper{}\taskBtimeout);\\
 \> \> \}\\
 \> \}\\
	\\
 \> {\bf do} \{ \\
 \> \> getMessage(\amper{}$m$);\\
 \> \> {\bf switch} ($m$.\emph{type}) \ \{\\
 \> \> \> TASK1;\\
 \> \> \> TASK2;\\
 \> \> \> TASK3;\\
 \> \> \}\\
 \> \} {\bf forever};
\end{tabbing}
\end{tt}

where tasks and actions are defined as follows:

\begin{tt}
\begin{tabbing}
aa\=TASK1 $\equiv$ \=aa\=aa\=aa\=aa\=aa \kill
 \> TASK1 $\equiv$ \> {\bf case} REPEAT\_TASK1:\\
 \> \> \> sendAll(I\_AM\_ALIVE);\\
 \> \> {\bf break;}
        \\
 \> TASK2 $\equiv$ \> {\bf case} REPEAT\_TASK2:\\
 \> \> \> $q =$ $m$.\emph{id};\\
 \> \> \> {\bf if} ($\outputp[q] \equiv$ TRUST)\\
 \> \> \> \> $\outputp[q]=$ SUSPECT;\\
 \> \> {\bf break;}
        \\
 \> TASK3 $\equiv$ \> {\bf case} I\_AM\_ALIVE:\\
 \> \> \> $q =$ $m$.\emph{sender};\\
 \> \> \> {\bf if} ($\outputp[q] \equiv$ SUSPECT) \ \{\\
 \> \> \> \> $\outputp[q]=$ TRUST;\\
 \> \> \> \> $\deltap(q) = \deltap(q) + 1$;\\
 \> \> \> \}\\
 \> \> {\bf break;}\\
	\\
 \> action\_Repeat\_Task1() \{\\
 \> \> {\bf message\_t} $m$;\\
 \> \> $m$.\emph{type} = REPEAT\_TASK1;\\
 \> \> Send($m$, $p$);\\
 \> \}\\
 \> action\_Repeat\_Task2({\bf timeout\_t} *$t$) \{\\
 \> \> {\bf message\_t} $m$;\\
 \> \> $m$.\emph{type} = REPEAT\_TASK2;\\
 \> \> $m$.\emph{id} = $t$->\emph{id};\\
 \> \> Send($m$, $p$);\\
 \> \}
\end{tabbing}
\end{tt}

We can draw the following observations:
\begin{itemize}
\item Our syntax is less abstract than the one adopted in the classical
formulation. Indeed we have deliberately chosen a syntax
very similar to that of programming languages such as C or C++.
Behind the lines, we assume also a similar semantics.
\item Our syntax is more strongly typed: we have deliberately chosen
to define (most of) the objects our code deals with.
\item We have systematically avoided set-wise operations such as union,
complement or membership by translating sets into arrays as, e.g., in
 $$\outputp \leftarrow \outputp \cup \{q\},$$
which we changed into
 $$\outputp[q] = \hbox{\tt PRESENT.}$$
\item We have systematically rewritten the abstract constructs {\tt repeat pe\-ri\-o\-di\-cal\-ly}
as one or more time-outs (depending on their multiplicity). Each of these time-out
has an associated action that sends one message to the client process, $p$. This means that 
\begin{enumerate}
\item time-related event ``it's time to send $p$-is-alive to all'' becomes event
``message {\tt REPEAT\_TASK1} has arrived.'' 
\item time-related events ``it's time to check whether $q$-is-alive has arrived''
becomes event ``message ({\tt REPEAT\_TASK2}, id=$q$) has arrived.''
\end{enumerate}
\item Due to the now homogeneous nature of the possible events (that now are all represented
by message arrivals) a single
process may manage those events through a multiple selection statement (a switch).
In other words, no coroutine is needed anymore.
\end{itemize}

Through the Literate Programming approach and a compliant
tool such as CWEB~\cite{KnLe93,Knuth92}
it is possible to further improve our reformulation. As well known, the CWEB tool allows
a pretty printable \TeX{} documentation
and a C file ready for compilation and testing to be produced from
a single source code.
In our experience this link between
these two contexts can be very beneficial: testing or even simply using the code
provides feedback on the specification of the algorithm, while the improved
specification may reduce the probability of design faults and in general
increase the quality of the code.

\begin{figure}[h]
\hskip-15pt\includegraphics[width=1.05\textwidth]{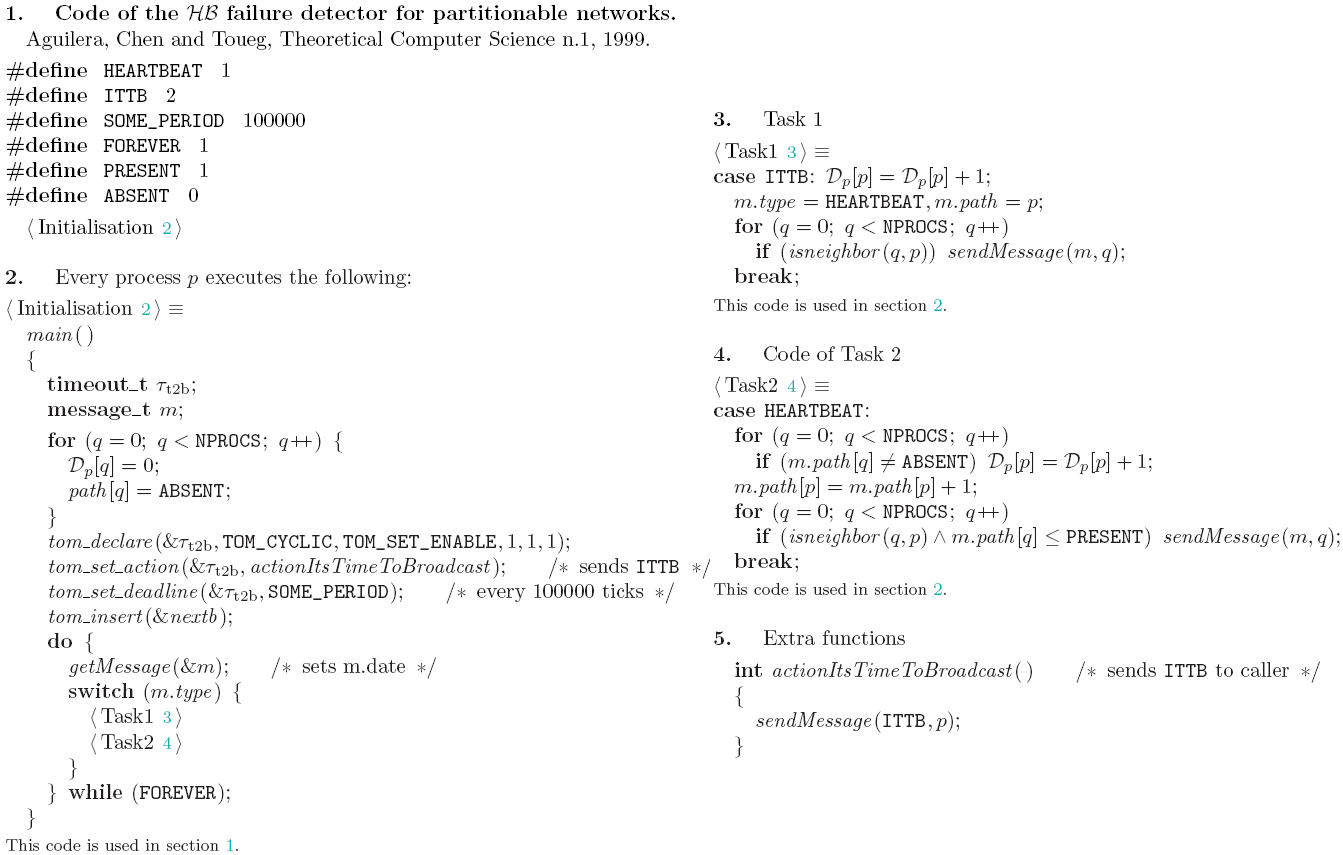}
\caption{Reformulation of the $\mathcal{HB}$ failure detector for partitionable networks~\cite{Aguilera99}.}
\label{f:Agui}
\end{figure}

\begin{figure}[h]
\hskip-15pt\includegraphics[width=1.05\textwidth]{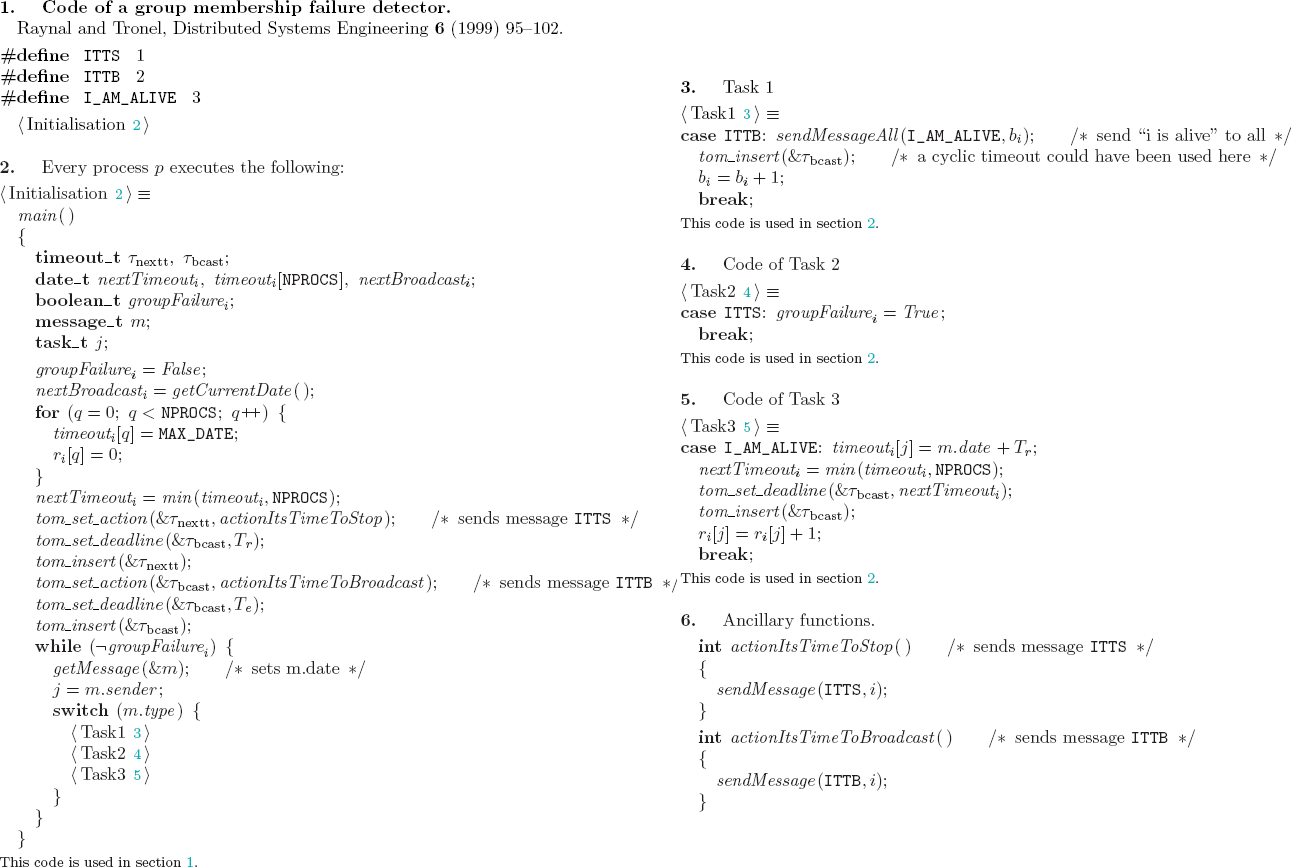}
\caption{Reformulation of the group membership failure detector~\cite{RaTr99}.}
\label{f:RaTr}
\end{figure}

Figure~\ref{f:Agui} and Figure~\ref{f:RaTr}
respectively show a reformulation for the $\mathcal{HB}$ failure detector for
partitionable networks~\cite{Aguilera99} and for the group membership failure detector~\cite{RaTr99}
produced with CWEB.
In those reformulations,
symbols such as $\tau$ and ${\mathcal D}_p$ are caught by CWEB and translated into legal C tokens
via its ``@f'' construct~\cite{KnLe93}. Note also that
the expression $m.\hbox{\em path}[q]\leq$\texttt{PRESENT} in Fig.~\ref{f:RaTr} means ``$q$ appears
at most once in \hbox{\emph{path}}''.
A full description of these protocols is out of the scope of this paper---for that we refer the
reader to the above cited articles. The focus here is mainly on the syntactical constructs
used in them and our reformulations, which
include simple translations for the syntactical
constructs in Table~\ref{t:syntax} in terms of our time-out API.
A case worth noting is that of the group membership failure detector: here
the authors mimic the availability of a cyclic time-out service but intrude its management
in their formulation. This management code can be avoided altogether using our approach.


\section{A development experience: the DIR net}\label{s:bb}
What we call ``DIR net''~\cite{DeFl09} is the distributed application at the core of the
software fault tolerance strategy realized through several European projects~\cite{DeFl09,BDDC99b}.
In this section we describe the DIR net and report on how we designed and developed it
by means of the TOM system.

The DIR net is a fault-tolerant network of
failure detectors connected to other peripheral error detectors
(called ``Dtools'' in what follows).
Objective of the DIR net is to ensure consistent fault tolerance
strategies throughout the system and play the role of a backbone
handling information to and from the Dtools~\cite{DeDL00d}.

The DIR net consists of four classes of components. Each processing node
in the system runs an instance of a so-called ``I'm Alive Task'' (IAT) plus
an instance of either a ``DIR Manager'' (\DIRM), or a ``DIR Agent'' (\DIRA), or
a ``DIR Backup Agent'' (\DIRB). A \DIRA{} gathers all error detection messages 
produced by the Dtools on the current processing node and forwards 
them to the \DIRM{} and the \DIRB's. A \DIRB{} is a \DIRA{} which also maintains its messages
into a database located in central memory. It is connected to \DIRM{} and to the other
\DIRB{}'s and is eligible for election as a \DIRM. A \DIRM{} is a special
case of \DIRB{}. Unique within the system, the \DIRM{} is the
one component responsible for running error recovery strategies---see~\cite{DeFl09}
for a description of the latter. Let us use DIR-$x$ to address any non-IAT component
(i.e. the \DIRM{}, or a \DIRB{}, or a \DIRA.)

An important design goal of the DIR net is that of being tolerant to physical and design faults,
both permanent or intermittent, affecting up to all but one \DIRB.
This is accomplished
also through a failure detection protocol that we are going to describe in the rest of
this section.

\subsection{The DIR net failure detection protocol}
Our protocol consists of a local part and a distributed part. Each of them is
realized through our TOM class.

\subsubsection{DIR net protocol: local component}
As we already mentioned, each processing node hosts a DIR-$x$ and an IAT. These two
components run a simple algorithm: they share a local Boolean variable, the ``I'm Alive Flag'' (IAF).
The DIR-$x$ has to periodically set the IAF to \texttt{TRUE} while the IAT has to check periodically
that this has indeed occurred and reverts IAF to \texttt{FALSE}. If the IAT finds the IAF set to
\texttt{FALSE} it broadcasts message \mTEIF{} (``this entity is faulty'').

The cyclic tasks mentioned above can be easily modeled via two time-outs, \tIAS{} and \tIAC, described
in Table~\ref{t:mIAS} and Table~\ref{t:tIAS} (TimeNow being the system function returning the current
value of the clock register.)

\begin{table*}
{\begin{tabular}{|l|l|l|c|}
   \hline
   \bf Time-out & \bf Caller & \bf Action                                     & \bf Cyclic? \\ \hline
    \tIAS       &  DIR-$x$   & On TimeNow + \dIAS{} do send $\mIAS$ to Caller & Yes\\
    \tIAC       & IAT        & On TimeNow + \dIAC{} do send $\mIAC$ to IAT    & Yes\\ \hline
  \end{tabular}}{ }
\caption{Description of messages \mIAS{} and \mIAC.\label{t:mIAS}}
\end{table*}

\begin{table*}
{\begin{tabular}{|l|l|l|l|}
   \hline
   \bf Message  & \bf Receiver & \bf Explanation       & \bf Action \\ \hline
     \mIAS      &  DIR-$x$     & Time to set IAF   & IAF $\leftarrow$ \texttt{TRUE}\\
     \mIAC      & IAT $k$      & Time to check IAF & \textbf{if} (IAF $\equiv$ \texttt{FALSE})
                                                            SendAll(\mTEIF, $k$)\\
                &              &                   & \textbf{else} IAF $\leftarrow$ \texttt{FALSE},\\ \hline
  \end{tabular}}
\caption{Description of time-outs \tIAS{} and \tIAC.\label{t:tIAS}}
\end{table*}

Note that the time-outs' alarm functions do not clear/set the
flag---doing so a hung DIR-$x$ would go undetected. On the contrary, those
functions trigger the transmission of messages that once
received by healthy components trigger the execution of the meant actions.

The following is 
a pseudo-code for the IAT{} algorithm:
\begin{tt}
\begin{center}
\begin{tabbing}
aa\=aa\=aa\=aa\=aa\=aa \kill
 \> \mbox{\emph{The IAT $k$ executes as follows}:}\\
        \\
 \> {\bf timeout\_t} \tIAC;\\
 \> {\bf msg\_t} \emph{activationMessage}, $m$;\\
 \\
 \> tom\_declare(\amper{}\tIAC, TOM\_CYCLIC,\\
 \> \> \> TOM\_SET\_ENABLE, IAT\_CLEAR\_TIMEOUT, 0, \dIAC);\\
 \> tom\_set\_action(\amper{}\tIAC, actionSend\mIAC);\\
 \> tom\_insert(\amper{}\tIAC);\\
 \\
 \> Receive(\emph{activationMessage});\\
 \\
 \> {\bf forever} \{\\
 \> \> Receive($m$);\\
 \> \> {\bf if} ($m$.\emph{type} $\equiv$ \mIAC) \\
 \> \> \> {\bf if} (IAF $\equiv$ \texttt{TRUE}) IAF $\leftarrow$ \texttt{FALSE}; \\
 \> \> \> {\bf else} SendAll(\mTEIF, $k$); delete\_timeout(\amper{}\tIAC);\\
 \> \}\\
 \\
 \> actionSend\mIAC() \{ Send(\mIAC, IAT $k$); \}
\end{tabbing}
\end{center}
\end{tt}

The time-out formulation of the IAT algorithm is given in next section.

\subsubsection{DIR net protocol: distributed component}\label{s:dirnet.dist}

\begin{figure}[h]
\centerline{\includegraphics[width=0.4\textwidth]{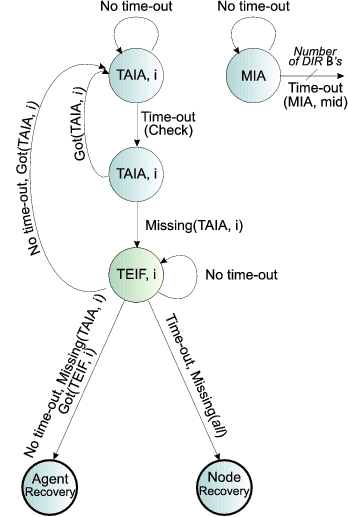}}
\caption{Algorithm of the \DIRM.}
\label{f:dadm}
\end{figure}

The resilience of the DIR net to crash faults comes from the \DIRM{} and
the \DIRB's running the following distributed algorithm of failure detection:

\paragraph{Algorithm \DIRM}\label{dirm}
Let us call {\tt mid} the node hosting the \DIRM{} and $b$ the number of processing nodes
that host a \DIRB.
The \DIRM{} has to send cyclically a \mMIA{} (``Manager-Is-Alive'') message
to all the \DIRB's each time time-out \tMIAM{} expires---this
is shown in the right side of Fig.~\ref{f:dadm}. Obviously this is a multiplicity $b$ ``repeat'' construct,
which can be easily managed through a cyclic time-out with an action that signals that a new cycle has begun.
In this case the action is ``send a message of type \mMIAM{} to the \DIRM.''

The manager also expects periodically a $(\mTAIA, \hbox{\small $i$})$ message (``This-Agent-Is-Alive'')
from each node where a \DIRB{} is expected to be running.
This is easily accomplished through a vector of
$(\tTAIAM, \hbox{\small $\mathbf i$})$ time-outs.
The left part of Fig.~\ref{f:dadm}
shows this for node $i$. When time-out $(\tTAIAM, \hbox{\small $p$})$ expires it means that
no $(\mTAIA, \hbox{\small $p$})$ message has been received within the current period. In this case
the \DIRM{} enters what we call a ``suspicion period''. During such period
the manager needs to distinguish the case of a late \DIRB{} from a crashed one.
This is done by inserting a non-cyclic time-out, namely $(\tTEIFM, \hbox{\small $p$})$.

During the suspicion period only one out of the following three events 
may occur:
\begin{enumerate}
\item A late  $(\mTAIA, \hbox{\small $p$})$ is received.
\item A  $(\mTEIF, \hbox{\small $p$})$ from IAT{} at node $p$ is received.
\item Nothing comes in and the time-out expires.
\end{enumerate}

In case {1.} we get out of the suspicion period, conclude that \DIRB{} at node $p$
was simply late and go back waiting for the next $(\mTAIA, \hbox{\small $p$})$.

It is the responsibility of the user to choose meaningful values for the time-outs'
deadlines. By ``meaningful'' we mean that those values should match the characteristics of the environment
and represent a good trade-off between the following two risks:

\begin{description}
\item[overshooting,] i.e., choosing too large values for the deadlines. 
This decreases the probability of false negatives (regarding a slow process as a failed process; this is
known as accuracy in failure detection terminology)
but increases the detection latency;
\item[undershooting,] namely under-dimensioning the deadlines. This may increase considerably
false negatives but reduces the detection latency of failed processes.
\end{description}

Under the hypotheses of properly chosen time-outs' deadlines, 
and that of a single, stable environment\footnote{We call an environment ``stable'' when
    it does not change drastically its characteristics except under erroneous and exceptional
    conditions. Single environments are typical of fixed (non-mobile) applications.},
the occurrences of late $(\mTAIA, \hbox{\small $p$})$ messages should be exceptional.
This event would translate in a false deduction uncovered in the next cycle.
Further late messages would postpone a correct assessment, but are considered as an unlikely
situation given the above hypotheses.
An alternative and better approach would be to track the changes in
the environment. For the case at hand this would mean that the time-outs' deadlines
should be adaptively adjusted. This could be possible, e.g., through an approach
such as in~\cite{DB07b}.

If {2.} is the case we assume the remote component has crashed though its node is still
working properly as the IAT{} on that node still gives signs of life. Consequently
we initiate an error recovery step. This includes sending a ``{\tt WAKEUP}''
message to the remote IAT{} so that it spawns another \DIRB{} on that node. 

In case {3.} we assume the entire node has crashed and initiate node recovery.

Underlying assumption of our algorithm is that the IAT{} is so simple that
if it fails then we can assume the whole node has failed.

\begin{figure}[h]
\centerline{\includegraphics[width=0.4\textwidth]{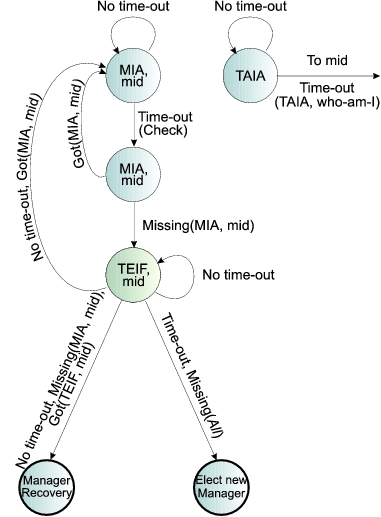}}
\caption{Algorithm \DIRB{}.}
\label{dadb}
\end{figure}

\paragraph{Algorithm \DIRB}\label{dirb}
This algorithm is also divided into two concurrent tasks. In the first one
\DIRB{} on node $i$ has to cyclically send 
$(\mTAIA, \hbox{\small $i$})$ messages to the manager, 
either in piggybacking or when time-out 
\tTAIAB{} expires.
This is represented in the right side of Fig.~\ref{dadb}.

The \DIRB{}'s in turn periodically expect a $\mMIA$ message from the \DIRM.
As evident when comparing Fig.~\ref{f:dadm} with Fig.~\ref{dadb},
the \DIRB{} algorithm is very similar to the one of the manager:
also \DIRB{}
enters a suspicion period when its manager does not appear to respond
quickly enough---this period is managed via time-out \tTEIFB,
the same way as in \DIRM. Also in this case we can get out of this state
in one out of three possible ways: either 
\begin{enumerate}
\item a late $(\mMIAB, \hbox{\small mid})$ is received, or
\item a $(\mTEIF, \hbox{\small mid})$ sent by the IAT{} at node mid is received, or
\item nothing comes in and the time-out expires.
\end{enumerate}
In case {1.} we get out of the suspicion period, conclude that the manager
was simply late, go back to normal state and start
waiting for the next $(\mMIA, \hbox{\small mid})$ message.
Also in this case, a wrong deduction shall be detected
in next cycles.
If 2. we conclude the manager has crashed though its node is still
working properly, as its IAT{} acted as expected. Consequently
we initiate a manager recovery phase structured similarly to
the \DIRB{} recovery step described in Sect.~\ref{dirm}.
In case 3. we assume the node of the manager has crashed,
elect a new manager among the \DIRB{}'s, and perform
a node recovery phase.

Table~\ref{t:tandm} summarizes the \DIRM{} and \DIRB{} algorithms.
\begin{table*}
{\vbox{\begin{tabular}{|l|l|l|c|}
   \hline
   \bf Time-out & \bf Caller & \bf Action                                           & \bf Cyclic? \\ \hline
    \tMIAM      & \DIRM      & Every \dMIAM{} do send \mMIAM{} to \DIRM             & Yes\\
   $\tTAIAM[i]$ & \DIRM      & Every \dTAIAM{} do send $(\mTAIAM,i)$ to \DIRM       & Yes\\
   $\tTEIFM[i]$ & \DIRM      & On TimeNow + \dTEIFM{} do send $(\mTEIFM,i)$ to \DIRM  & No\\
    \tTAIAB     & \DIRB{} $j$& Every \dTAIAB{} do send \mTAIAB{} to \DIRB{} $j$     & Yes\\
    \tMIAB      & \DIRB{} $j$& Every \dMIAB{} do send \mMIAB{} to \DIRB{} $j$       & Yes\\
    \tTEIFB     & \DIRB{} $j$& On TimeNow + \dTEIFB{} do send $\mTEIFB$ to \DIRB{} $j$& No \\ \hline
  \end{tabular}\\
  \begin{tabular}{|l|l|l|l|}
   \hline
   \bf Message  & \bf Receiver & \bf Explanation                & \bf Action \\ \hline
   $(\mTAIA,i)$ & \DIRM        & \DIRB{} $i$ is OK              & (Re-)Insert or renew $\tTAIAM[i]$ \\
     \mMIAM     & \DIRM        & A new heartbeat is required    & Send \mMIA{} to all \DIRB's \\
     \mTAIAM    & \DIRM        & Possibly \DIRB{} $i$ is not OK & Delete $\tTAIAM[i]$, insert $\tTEIFM[i]$\\
   $(\mTEIF,i)$ & \DIRM        & \DIRB{} $i$ crashed        & Declare \DIRB{} $i$ crashed\\ 
   $(\mTEIFM,i)$& \DIRM        & Node $i$ crashed           & Declare node $i$ crashed\\ 
     \mMIA      & \DIRB{} $j$  & \DIRM{} is OK                  & Renew \tMIAB\\
     \mTAIAB    & \DIRB{} $j$  & A new heartbeat is required    & Send $(\mTAIA,j)$ to \DIRM\\
     \mMIAB     & \DIRB{} $j$  & Possibly \DIRM{} is not OK     & Delete \tMIAB, insert \tTEIFB\\
     \mTEIF     & \DIRB{} $j$  & \DIRM{} crashed            & Declare \DIRM{} crashed\\
     \mTEIFB    & \DIRB{} $j$  & \DIRM{}'s node crashed     & Declare \DIRM's node crashed\\ \hline
  \end{tabular}}}
\caption{Time-outs and messages of \DIRM{} and \DIRB{}.\label{t:tandm}}
\end{table*}

We have developed the DIR net using the Windows TIRAN libraries~\cite{BDDC99b} and the CWEB system
of structured documentation.

\subsection{Special services}

\subsubsection{Configuration}
\begin{figure}[h]
\centerline{\includegraphics[width=0.7\textwidth]{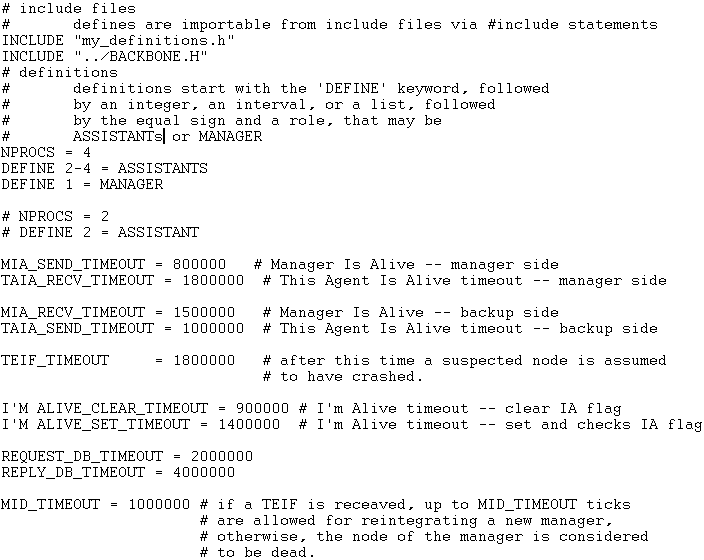}}
\caption{Excerpt from the configuration script of the DIR net.}
\label{f:xxx4}
\end{figure}

\begin{figure}[h]
\centerline{\includegraphics[width=0.7\textwidth]{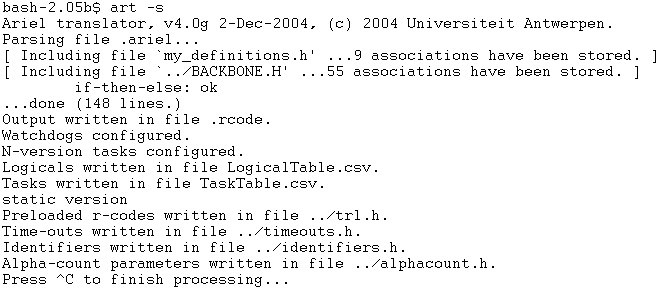}}
\caption{Configuration tool of the DIR net.}
\label{f:xxx5}
\end{figure}

The management of a large number of time-outs may be an error prone task. To simplify it,
we designed a simple configuration language.
Figure~\ref{f:xxx4} shows an example of configuration script 
to specify the structure of the DIR net
(in this case, a four node system with three \DIRB{}'s deployed on nodes 1--3 and the \DIRM{}
on node 0) and of its time-outs.
A translator produces the C header files to properly initialize an instance of the DIR net
(see Fig.~\ref{f:xxx5}).

\subsubsection{Fault injection}
\begin{figure}[h]
\includegraphics[width=0.82\textwidth]{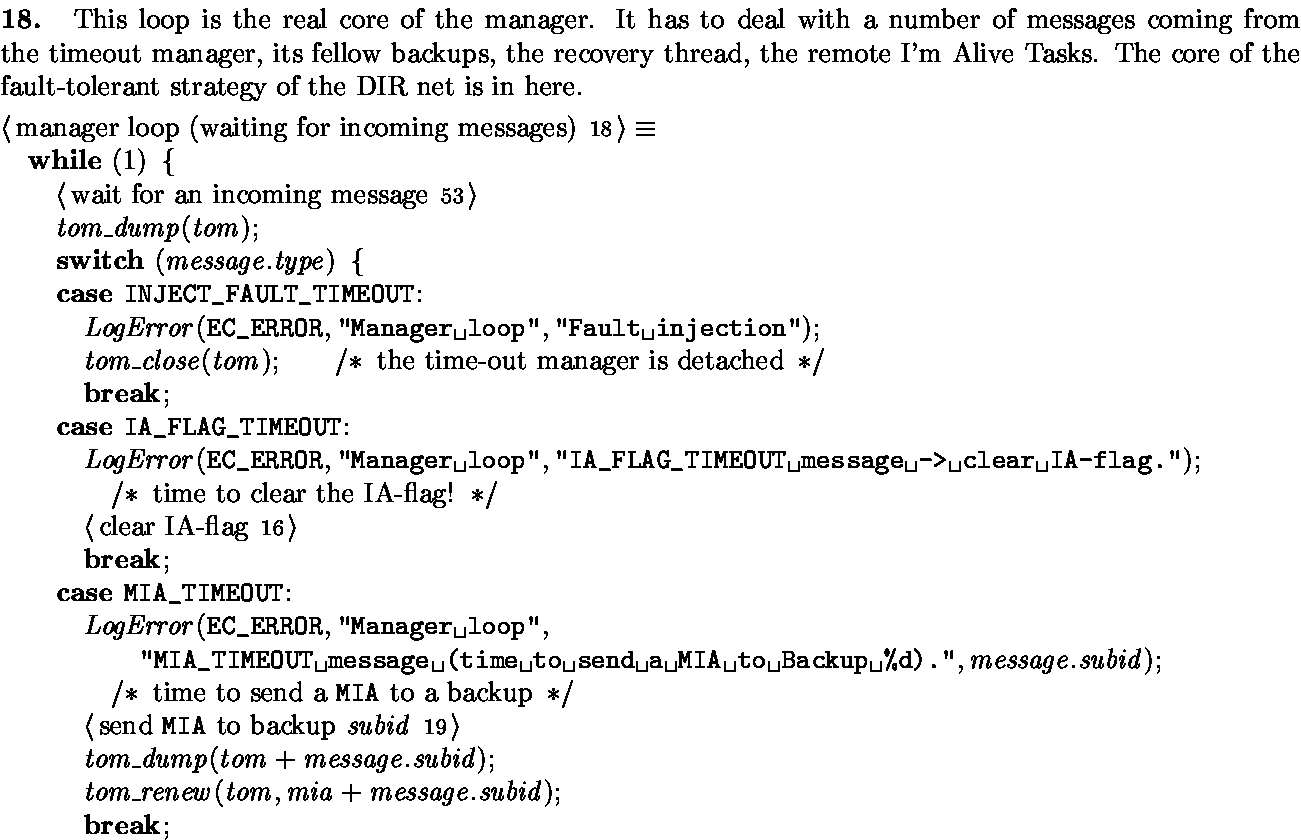}
\caption{Excerpt from the CWEB source of the DIR net.}
\label{f:mainloop}
\end{figure}

Time-outs may also be used to specify fault injection actions with fixed or pseudo-random deadlines.
In the DIR net this is done as follows. First we define the time-out:

\begin{verbatim}
#ifdef INJECT
 tom_declare(&inject, TOM_NON_CYCLIC, TOM_SET_ENABLE,
             INJECT_FAULT_TIMEOUT, i, INJECT_FAULT_DEADLINE);
 tom_insert(tom, &inject);
#endif
\end{verbatim}

The alarm of this time-out sends the local DIR-$x$ a message of type ``\texttt{IN\-JE\-CT\_FA\-ULT\_TIME\-OUT}''.
Figure~\ref{f:mainloop} shows an excerpt from the actual main loop of the \DIRM{} in which 
this message is processed.

\subsubsection{Fault tolerance}
A service such as TOM is indeed a single-point-of-failure in that 
a failed TOM in the DIR net would result in all components being unable to
perform their failure detection protocols.
Such a case would be indistinguishable from that of a crashed node by the other DIR net components.
As well known from, e.g., \cite{IBR96}, a single design fault in TOM's implementation could bring
the system to a global failure. Nevertheless, the isolation of a \emph{service\/} for time-out management
may pave the way for a cost-effective adoption of multiple-version software fault tolerance techniques~\cite{Lyu98b}
such as the well known recovery block~\cite{RaXu95}, or $N$-version programming~\cite{Avi95}.
Another possibility would be to use the DIR net algorithm to tolerate faults in TOM.
No such technique has been adopted in the current implementation of TOM.
Other factors, such as congestion or malicious attacks might introduce performance
failures that would impact on all modules that depend on TOM to perform their time-based
processing~\cite{DeFl06}.


\section{Conclusions}\label{s:end}
We have introduced a tentative \emph{lingua franca\/} for the expression of failure detection protocols.
TOM has the advantages of being simple, elegant and not ambiguous. Obvious are the
many positive relapses that would come from the adoption of a standard, semi-formal representation with respect to
the current Babel of informal descriptions---easier acquisition of insight, faster verification, and greater ability
to rapid-prototype software systems. The availability of a tool such as TOM is also one of
the requirements of the timed-asynchronous system model~\cite{CrFe99}.

Given the current lack of a network service for failure detection, the availability of standard methods
to express failure detectors in the application layer is an important asset: a tool like the one
described in this paper isolates and crystallizes a part of the complexity required to express failure detection protocols. 
This complexity may become transparent of the designer, with tangible savings
in terms of development times and costs, if more efforts will be devoted to
time-outs configuration and automatic adjustments through adaptive 
approaches such as the one described in~\cite{DB07b}. Such optimizations will be
the subject of future research.
Future plans also include to port our system to AspectJ~\cite{aspectj} so as to further enhance
programmability and separation of design concerns.

As a final remark we would like to point out how, at the core of our design choices,
is the selection of C and literate programming, which proved to be invaluable tools
to reach our design goals. Nevertheless we must point out how these choices
may turn into intrinsic limitations for the expressiveness of the resulting language. 
In particular, they enforce a syntactical and semantic structure, that of the C programming language,
which may be regarded as a limitation by those researchers who are not accustomed to that language.
At the same time we would like to remark also
that those very choices allow us a straightforward translation of our
constructs into a language like Promela~\cite{Hol91}, which resembles very much a C language
augmented with Hoare's CSP~\cite{Hoa78}.
Accordingly, our future work in this framework shall include the adoption of the Promela
extension of Prof. Bo{\v s}na{\v c}ki, which allows the verification of concurrent systems that depend
on timing parameters~\cite{BoDa98}. Interestingly enough, this version of Promela 
includes new objects, called discrete time countdown timers, which are basically equivalent to our 
non-cyclic time-outs. Our goal is to come up with a tool that generates
from the same literate programming source (1) a pretty printout in \TeX,
(2) C code ready to be compiled and run, and (3) Promela code to verify some properties of the protocol.

\section*{Acknowledgment}
We acknowledge the work by Alessandro Sanchini, who developed
the communication library used by our tool, and the many and valuable comments
of our reviewers.

\newpage


\vfill\eject

\vfill\eject

\vfill\eject

\vfill\eject

\vfill\eject

\vfill\eject

\vfill\eject

\vfill\eject

\vfill\eject

\end{document}